\documentclass[12pt]{article}
\newcommand\be{\begin{equation}}
\newcommand\ee{\end{equation}}
\newcommand\ba{\begin{eqnarray}}
\newcommand\ea{\end{eqnarray}}
\newcommand\eq{\begin{equation}}           
\newcommand\en{\end{equation}}

\def\ga{\mathrel{\rlap {\raise.5ex\hbox{$>$}}
{\lower.5ex\hbox{$\sim$}}}}
\def\la{\mathrel{\rlap{\raise.5ex\hbox{$<$}}
{\lower.5ex\hbox{$\sim$}}}}

\input epsf
\usepackage{amssymb}
 \usepackage{epsfig,cite}
\include{epsf}

\usepackage{subfigure}

\begin{document}
\title{
{\hfill  \small MCTP-09-47,FTPI-MINN-09/34, UMN-TH-2814/09  \\ ~\\~\\}
Heavy Right-Handed Neutrinos and Dark Matter in the $\nu$CMSSM}

\author{Kenji Kadota$^1$ and Keith A. Olive$^2$ \\
{\em \small 
$^1$ Michigan Center for Theoretical Physics, University of Michigan, Ann Arbor, MI 48109} \\
 {\em \small
$^2$ William I. Fine Theoretical Physics Institute, University of Minnesota, Minneapolis, MN 55455}
}
\maketitle   

\begin{abstract}
We perform a systematic study of the effects of the type-I seesaw mechanism on the dark matter abundance 
in the constrained supersymmetric standard model (CMSSM) which 
includes three right-handed neutrinos (the $\nu$CMSSM).
For large values of $m_0,m_{1/2}$, we exploit the effects of large neutrino Yukawa couplings on the renormalization group
(RG) evolution of the up-type Higgs. 
In particular, we show that the focus point scale can greatly exceed the 
electroweak scale resulting in the absence of a focus point region for which the relic density of 
neutralinos is within the range determined by WMAP. 
We also discuss the effects of the right-handed neutrinos on the so-called funnel region,
where the relic density is controlled by s-channel annihilations through a heavy Higgs.
For small values of $m_0,m_{1/2}$, we discuss the possibility of 
sneutrino coannihilation regions with an emphasis on the 
suppression of the left-handed slepton doublet masses due to the neutrino Yukawa coupling.
We consider two types of toy models consistent with either the normal or inverted 
hierarchy of neutrino masses.
\end{abstract}

\setcounter{footnote}{0} 
\setcounter{page}{1}
\setcounter{section}{0} \setcounter{subsection}{0}
\setcounter{subsubsection}{0}

\section{Introduction}
The seesaw mechanism offers a compelling explanation for the tiny masses of the left-handed neutrinos 
by introducing heavy gauge-singlet right-handed neutrinos which can alleviate 
the fine-tunings of the neutrino Yukawa couplings \cite{seesaw}. 
If the right-handed neutrino masses are close to the grand unification (GUT) scale, one typically
finds neutrino Yukawa couplings of order unity.
These heavy degrees of freedom may also influence such effects as lepton flavor violation
\cite{borzu,hisa,casa3,Lavignac:2001vp,Ellis:2001xt,bm,de,iba,hir} and the mass/coupling spectrum due to changes in the RG(renormalization group) evolution \cite{casa,how4,baer2,bla,hitoshi2,gordy3,hints,kov}. 
In this paper, we implement the seesaw mechanism in the context of the constrained minimal supersymmetric model (CMSSM) by adding three right-handed neutrinos $N_i$. We study the consequences of the near-GUT-scale right-handed neutrinos in such a model (hereafter called the $\nu$CMSSM) on the thermal dark matter relic abundance which is calculated around the electroweak scale.  
The effects of the seesaw mechanisms on the 
dark matter abundance has been studied for several GUT and seesaw schemes \cite{petcov,cali,barg,gom,est,barg2} indicating signatures different from those in conventional CMSSM scenarios \cite{funnel,cmssm,efgosi,analytic,dm,eoss,cmssmwmap,hb}, such as the sneutrino coannihilation regions \cite{kov}.     

Here, we present a systematic study of the effects of a general type-I seesaw without any additional constraints from a specific GUT model which may require Yukawa coupling unification.  
We show in detail the behavior of two of the standard CMSSM regions where the 
WMAP relic density \cite{wmap} is achieved, namely the funnel region, where the mass of the lightest supersymmetric particle (LSP), the neutralino, is roughly half the mass of the pseudo-scalar Higgs boson \cite{funnel,efgosi}, and the focus point region which borders the region where radiative electroweak  symmetry breaking is no longer possible \cite{hb,analytic,dm}.
In the case of the latter, the
 behavior of the Higgs mass drastically changes in the
$\nu$CMSSM because of the change in the focus point scale from the RG evolution of  $m^2_{H_u}$. 
As we will see, the 
focus point region can even completely disappear once the energy scale of the focus point is sufficiently 
greater than the weak scale. As a result the WMAP strip associated with the focus point can cease 
to exist once the right-handed neutrino mass scale, $M_N$, is sufficiently large.  The funnel region is 
also affected (though not as dramatically) as the pseudo-scalar mass is also affected by
the choice of $M_N$.

We also detail changes in the stau coannihilation region which occurs in the CMSSM near the 
line of degeneracy between the neutralino and lighter stau masses. As we have shown
previously \cite{kov}, the introduction of a right handed neutrino can affect the running of the left-handed
sleptons so that they become lighter than the more common next-to-LSP, the predominantly right-handed stau.  This has the effect that the sneutrino may well become the NLSP.  Here, we also extend 
that work to include the effects of all three right-handed neutrino states.

We review the essential features of the $\nu$CMSSM in \S\ref{nucmssm}. 
In \S \ref{foc}, we consider the effect of the right-handed neutrinos on the location of the focus point regions in the $\nu$CMSSM. 
We preform an analogous analysis for the Higgs funnel regions in \S \ref{higfun}. The sneutrino coannihilation regions with small $m_0,m_{1/2}$  are studied in \S\ref{invert}, followed by the conclusions in \S\ref{conc}.

\section{$\nu$CMSSM}
\label{nucmssm}
For simplicity, we do not consider flavor mixings in the neutrino sector, and as such, 
in the basis where the heavy right-handed neutrino mass 
matrix is diagonal\footnote{We can always make the right-handed neutrino mass and lepton Yukawa coupling matrices diagonal by 
unitary transformations of $N^c,L$ and $E^c$. The diagonal form of 
the neutrino Yukawa coupling matrix in such a basis is the simplification we make here.}
our $\nu$CMSSM superpotential can be written as
\ba
W=W_{CMSSM}+y_{N_i} N^c_i L_i H_u + \frac12 M_{N_i} N^c_iN^c_i  .
\ea
We assume that all soft SUSY breaking parameters including 
the soft SUSY breaking sneutrino masses, 
$m_{N_i}$, are universal at the GUT scale with value $m_0$. 
In addition, we also assume the SUSY breaking trilinear couplings
including $A_{N_i}$ are universal at the GUT scale. 
As in the CMSSM, we take universal gaugino masses.
Then, in addition to the conventional CMSSM parameter set
$m_0,m_{1/2},A_0,\tan \beta,sgn(\mu)$, the 
$\nu$CMSSM has the following additional parameters 
\ba
M_{N_1}(Q_{GUT}),M_{N_2}(Q_{GUT}),M_{N_3}(Q_{GUT}),
m_{\nu_1}(Q_{m_Z}),m_{\nu_2}(Q_{m_Z}),m_{\nu_3}(Q_{m_Z}) ,
\ea
where the heavy right-handed neutrino masses $M_{N_i}$ are specified at the 
GUT scale $Q_{GUT}$, while the light left-handed neutrino masses are 
specified at the weak scale, $Q_{m_Z}$. 
These boundary conditions are sufficient to determine the neutrino Yukawa couplings $y_{N_i}$ 
as well. In this analysis, we restrict the right-handed neutrino masses to lie below the GUT scale $M_{N_i}<Q_{GUT}$. We evolve the full two-loop renormalization group equations (RGEs) in 
the $\nu$CMSSM. 
Each of the right-handed neutrinos decouples at $M_{N_i}$ which itself 
runs from the GUT scale to $M_{N_i}$. Below the energy scale $M_{N_i}$, 
$N_i$ is integrated out and the effective Lagrangian includes the residual 
higher order operators suppressed by $M_{N_i}$. Because we are interested in
 the value of $M_{N_i}$ relatively close to the GUT scale in the type-I seesaw scheme, those non-renormalizable operators 
do not affect the mass spectrum at the low energy scale other than the left-handed neutrino masses which receive a dominant contribution from the dimension five operator
\ba
L_5  \ni -\kappa (L H_u)(L H_u)
\ea 
Hence, we keep the running of $\kappa$ in our 
full two-loop RGEs and $\kappa \langle H_u \rangle^2 $ 
is matched to $m_{\nu}(Q_{m_Z})$ by our electroweak scale boundary conditions 
 \cite{ant,ant2}.    

Even though our numerical analysis was performed including full two-loop RGEs, it will be useful 
to use the one-loop RGE expressions for a qualitative discussion. 
The relevant RGEs are ($t \equiv logQ $)
\ba
\frac{d}{dt}m_{H_u}^2&=& \frac{1}{16 \pi^2}\left(Tr\left[
6(m_{H_u}^2+{\bf m}^2_Q){\bf y}_u^{\dagger}{\bf y}_u
+6{\bf y}_u^{\dagger}{\bf m}^2_u{\bf y}_u
+6 {\bf h}_u^{\dagger}{\bf h}_u
+2(m_{H_u}^2+{\bf m}^2_L){\bf y}_N^{\dagger}{\bf y}_N \right. \right. \nonumber \\
&& \left. \left. +2{\bf y}_N^{\dagger}{\bf m}^2_N{\bf y}_N
+2 {\bf h}_N^{\dagger}{\bf h}_N
\right]
-\frac{6}{5}g_1^2 M_1^2-6g_2^2 M_2^2+\frac{3}{5}g_1^2 S \right)+... \nonumber 
\\
\frac{d}{dt}{\bf m}_{L}^2&=&\frac{1}{16 \pi^2}\left(
(2m_{H_d}^2+{\bf m}^2_L){\bf y}_e^{\dagger}{\bf y}_e
+2{\bf y}_e^{\dagger}{\bf m}^2_E{\bf y}_e
+{\bf y}_e^{\dagger}{\bf y}_e{\bf m}^2_L
+2 {\bf h}_e^{\dagger}{\bf h}_e
+2{\bf y}_N^{\dagger}{\bf m}^2_N{\bf y}_N \right.  \nonumber \\
&&\left.
+{\bf y}_N^{\dagger}{\bf y}_N{\bf m}^2_L
+2 {\bf h}_N^{\dagger}{\bf h}_N
+(2m_{H_u}^2 +{\bf m}^2_L){\bf y}_N^{\dagger}{\bf y}_N
-\frac65g_1^2 M_1^2-6g_2^2 M_2^2-\frac{3}{5}g_1^2 S\right)+... \nonumber \\
\frac{d}{dt}{\bf y}_{u}&=&
\frac{{\bf y}_{u}}{16 \pi^2}\left(
Tr\left[3{\bf y}_u {\bf y}_u^{\dagger}+{\bf y}_N {\bf y}_N^{\dagger} \right]
+3{\bf y}_u^{\dagger}{\bf y}_u+{\bf y}_d^{\dagger}{\bf y}_d
-\frac{13}{15}g_1^2 -3g_2^2 -\frac{16}{3}g_3^2
\right)+...     \nonumber      \\
S&=&Tr({\bf m}_{ Q}^2 + {\bf m}_{ D}^2 -2 {\bf m}^2_{ U}
-{\bf m}_{ L}^2 +{\bf m}_{E}^2)
+ {m}^2_{H_u}-{m}^2_{H_d}
\label{mhurge} 
\ea
where, for the models to be discussed below, $h_{ij}=A_{ij}y_{ij}\delta_{ij}$. In the CMSSM as well as in the $\nu$CMSSM, $S=0$ at the GUT scale due to the universality of the soft scalar masses. Deviations from $S=0$ are due to the RG evolution at the two-loop level, hence, $S$ does not play a significant role in our study.
In the following sections, the RG effects of large neutrino Yukawa couplings on $m_{H_u}^2$ 
are studied in the focus point and Higgs funnel regions whereas those effects on 
the slepton doublets ${\bf m}^2_{L}$ are studied in the sneutrino/stau coannihilation
regions.

We leave the question of flavor mixings and the CP violating phases for future work and we 
present here the analysis for two simple toy models to illustrate the effects of the heavy right-handed neutrinos on the thermal dark matter relic abundance. The first model includes only one right-handed neutrino in the third generation $N_3$ consistent with a 
normal mass hierarchy $m_{\nu_3}\gg m_{\nu_1}=m_{\nu_2}$. To see the effects of the 
additional generations with multiple right-handed neutrinos, our second model includes two heavy right-handed neutrinos in the first two generations $N_1,N_2$ leading to an inverted mass hierarchy spectrum $m_{\nu_3}\ll m_{\nu_1}=m_{\nu_2}$. 

 We start our discussion with the first  model including only $N_3$ 
and and concentrate on large values of $m_0,m_{1/2}$ in sections \S \ref{foc} and \S \ref{higfun}. In section \S\ref{invert}, we go on to discuss our second model including two right-handed neutrinos in the first two generations $N_1,N_2$ at relatively small $m_0,m_{1/2}$ relevant  for the sneutrino/stau coannihilation regions.

\section{The focus point region}
\label{foc}
\begin{figure}
             \centering

             {
                 \includegraphics[width=0.47\textwidth]{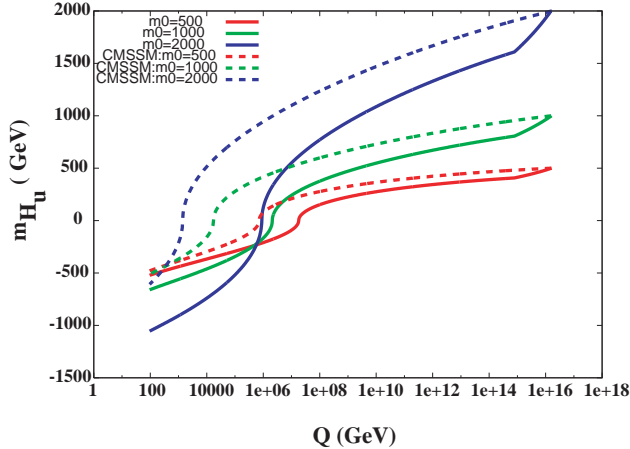}
             }
             
\caption{The RG evolutions of $m_{H_u}$ with different values of $m_0$ for the 
CMSSM (dashed) and the $\nu$CMSSM with $M_{N_3}=10^{15}$GeV and $m_{\nu_3}=0.05$eV (solid).  We have set $m_{1/2}=300$~GeV, $\tan \beta=10, A_0=0,$ and $\mu>0$.}
     \label{focus1} 
     \end{figure}

When the Higgsino mass parameter, $\mu$, becomes small or comparable to $m_{1/2}$,
the lightest neutralino (which is typically the LSP) has 
an increasingly large Higgsino component and neutralino 
annihilations become dominated by $W^+ W^-$ final states through 
t-channel chargino exchange.  The result of the enhanced annihilation cross section
is a thermal neutralino relic abundance in the desired WMAP (95\% CL) range \cite{wmap}
\ba
0.0975 < \Omega_\chi h^2 < 0.1223
\label{wmap}
\ea 
in a region (large $m_0$) where nominally the relic density is expected to be large.
This region is close to the limit (at large $m_0$) where there cease to be solutions
to the Higgs potential minimization conditions corresponding to radiative electroweak 
symmetry breaking.  In the CMSSM, the RGEs exhibit a focusing property leading to 
what is often referred to as
the focus point or hyperbolic branch regions \cite{analytic,hb,dm}. 
The RG trajectories of $m^2_{H_u} \sim -|\mu|^2$ for different values of $m_0$ (and $\tan \beta$) meet (or focus) near the weak scale \cite{analytic}. 
The focusing of  the RG trajectories of $m^2_{H_u}$  also occurs in the $\nu$CMSSM, as the
RGEs for the gaugino masses and trilinear $A$ parameters remain independent of the 
scalar masses at the one-loop level 
(while those for the scalar masses depend on the scalar masses, gaugino masses and $A$ parameters). 
In the $\nu$CMSSM, however, the focus point scale where RG trajectories meet 
is not necessarily found at the weak scale due to large neutrino Yukawa contributions. 
The RG evolutions of $m_{H_u}$ in the CMSSM and $\nu$CMSSM are illustrated in Fig. \ref{focus1} 
for different values of $m_0$, where the sign of $m_{H_u}$ indicates the sign of $m^2_{H_u}$. 
We use a top mass $m_t=173.1$ GeV \cite{top} and a bottom mass in the minimal subtraction (MS) scheme $m_b(m_b)^{MS}=4.25$ GeV in our analysis.\footnote{See, for instance, Refs. \cite{efgosi,analytic,mtvary,focusvary} for the effects of the variations of $m_t,m_b$ within the experimentally allowed ranges on the dark matter abundances, which however do not affect our discussions in this paper.}

Before presenting the numerical results for 
the dark matter abundance calculations, we briefly examine the RG evolution
of $m^2_{H_u}$ to illustrate the properties of the focus point regions which yield the 
observed thermal relic abundance.
The tree-level minimization of the Higgs potential gives 
\ba
|\mu|^2=\frac{m_{H_d}^2-m^2_{H_u}\tan^2 \beta}{\tan^2 \beta-1}-\frac{M^2_Z}{2}
\ea
This allows one to estimate   $\mu^2\approx -m^2_{H_u}-\frac12 m_Z^2$ for  moderate to large values of  $\tan \beta$ (we hereafter assume $\mu > 0$ in our analysis as favored by measurements of  $b\rightarrow s \gamma$ and the anomalous magnetic moment of the muon).
The RGE of $m^2_{H_u}$, which gives us an estimate for $\mu$ at the weak scale, 
is given in Eq. (\ref{mhurge}). In the CMSSM (or when $y_N = 0$), we see the RG evolution of $m^2_{H_u}$ is sensitive to the top Yukawa coupling \cite{ytop,nuhm}. In fact, in the CMSSM, the focus point scale $Q_{FP}$ is determined by the value of the top quark Yukawa coupling 
evaluated at the focus point and the (entire RG trajectories of) gauge couplings \cite{analytic}. 
The focus point scale of 
$m_{H_u}^2$ in the CMSSM turns out to be the weak scale for the measured value of top quark mass with a small dependence on moderate to large values of $\tan \beta$ (this can be inferred from the top Yukawa coupling  $y_t\propto 1/\sin \beta \propto \mbox{const}+{\cal O}(\tan^{-2} \beta)$). 
This behavior is seen in Fig. \ref{focus1} which shows the CMSSM running of $m^2_{H_u}$
for three values of $m_0$.  As one can clearly see, the three dashed curves corresponding to the CMSSM focus at a weak scale value for $Q$. 

In the $\nu$CMSSM, the effects of the neutrino Yukawa couplings on the RGEs of $m^2_{H_u}$ and $y_t$ can be significant as seen in Eq. (\ref{mhurge}), so that the focus point scale of $m^2_{H_u}$ can well exceed the weak scale.  This is seen in Fig. \ref{focus1} which shows the three solid curves
of the $\nu$CMSSM focusing at the scale $\sim 10^6$ GeV where $y_{N_3}(Q_{GUT})\sim 2$. Consequently, and
contrary to CMSSM, the insensitivity of $m_{H_u}^2$ evaluated at the weak scale to $m_0$ and $\tan \beta$ does not necessarily hold in the $\nu$CMSSM.\footnote{Once the heavy right-handed neutrino is integrated out at $Q=M_N$, the analytical estimation of the focus point analogous to CMSSM can be applied to the $\nu$CMSSM for $Q \leq M_N$ with 
non-universal scalar mass values at $Q=M_N$. Hence, as in the case of the CMSSM with non-universal boundary conditions, the focus point can still be found at the weak scale  in the
$\nu$CMSSM if one can tune the model parameters to realize the particular ratios of those scalar mass values at $Q=M_N$ \cite{analytic}.} The positions and existence of the focus point regions (at large $m_0$ where the relic density is sufficiently small) change due to this shift in the focus point scale. 
When the focus point scale is below the 
energy scale relevant for the neutralino annihilations at which $\mu$ is evaluated, we can decrease $\mu \sim \sqrt{-m^2_{H_u}}$  by choosing a larger value for  $m_0$ (see CMSSM curves in Fig. \ref{focus1}).  Recall that at small $\mu$,  neutralino annihilation is dominated by the Higgsino component  and can yield an acceptable relic density.
As one can also see from Fig. \ref{focus1}, when the focus point scale becomes large (as depicted in
the figure for the $\nu$CMSSM), the weak scale value of $m^2_{H_u}$ becomes sensitive to $m_0$
and is driven to large negative values when $m_0$ is large. 
In this case an increase of 
$m_0$ (with other parameters fixed) increases $\mu$ and the Higgsino component
of the LSP is diminished and no region of suitable relic density can be found.

\begin{figure}[htb!]
         \begin{center}
                  \includegraphics[width=0.47\textwidth]{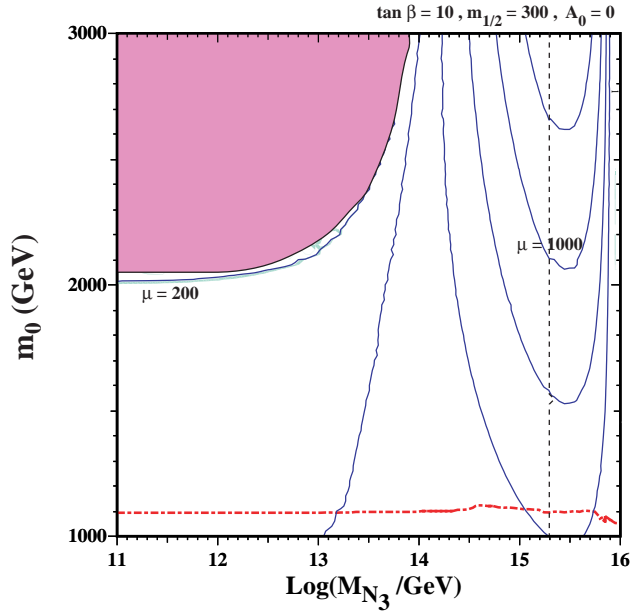}
            \end{center}
\caption{The ($m_0,M_{N_3}$) plane for fixed 
$m_{1/2}=300$ GeV, $\tan \beta =10, A_0=0$, and $m_{\nu_3}=0.05$ eV.
The (pink) shaded region shows the portion of parameter space
where the radiative electroweak symmetry breaking conditions can not be satisfied.
The focus point region (turquoise) 
compatible with Eq. (\ref{wmap}) runs along this boundary nearly coinciding
with the contour for $\mu = 200$ GeV.  Other contours of constant $\mu$ are also shown.
The (red) dot-dashed curve shows the contour of $m_h = 114.4$ GeV.  Below this curve, the 
Higgs mass is below the LEP limit.  
To the right of the vertical dashed line, ${y^2_{N_3}}/{16\pi^2}>1/10$ and our perturbative
expansion becomes suspect.}
     \label{focusMnm0} 
     \end{figure}

 We also note that, for a given value of $m_0$, the value of $m_{H_u}(Q)$ is smaller in the
$\nu$CMSSM  than in the CMSSM, as can be inferred from the 
form of RGE of $m_{H_u}^2$ in Eq. (\ref{mhurge}) due to the additional terms involving $y_N^2$.
The behavior of these mass parameters is also illustrated in Fig. \ref{focusMnm0} where we also show the shift/disappearance of the boundary where radiative electroweak symmetry is no longer possible.
Inside the (pink) shaded region, the Higgs potential minimization conditions yield $\mu^2 < 0$. Along
the boundary of that region $\mu=0$, and close to the boundary, we show a
contour of constant $\mu = 200$ GeV.  Almost coinciding with that contour is
the very thin WMAP focus point region (shaded turquoise)
where the relic density lies in the range given in Eq. (\ref{wmap}).  Also shown are
contours of constant $\mu$ at 200 GeV intervals. As one can see, as $M_{N_3}$ is increased, the
focus point region shifts to increasing $m_0$ and is absent at $M_{N_3} \sim 10^{14}$ GeV
(which corresponds to $y_{N_3}(Q_{GUT})\sim 0.6$) 
at which point there is no longer any sensitivity of the value of $\mu$ to $m_0$.
This corresponds to the coincidence of the focus point scale with the 
scale at which $\mu$ is evaluated.
In this figure, we have fixed $m_{1/2} = 300$ GeV, $\tan \beta = 10$ and $A_0 = 0$. 
For $M_{N_3} \la 10^{14}$ GeV, the focus point energy scale, $Q_{FP}$ is low and increasing 
$m_0$ helps making $m_{H_u}^2$ less negative which reduces $\mu$ allowing for the presence of  
the focus point region. 
For larger $M_{N_3}$, the opposite behavior occurs and increasing $m_0$ increases $\mu$. 

Fig. \ref{focusMnm0} also shows the region where 
the two-loop contributions become large.  To the right of the vertical dashed line at 
$M_{N_3}\sim 2 \times 10^{15}$ GeV, ${y^2_{N_3}}/{16\pi^2}>{1}/{10}$ (where the dominant
two loop contribution is roughly 1/4 of the 1-loop contribution).  At larger values of $M_{N_3}$,
our perturbative calculation results could become unreliable \cite{casa3,bm,nume}.
To the left of this line, we
see that $\mu$ increases with $M_{N_3}$ for fixed $m_0$. 
Finally, in the region below the (red) dot-dashed curve, $m_h < 114.4$ GeV and 
fails the LEP limit \cite{LEPsusy} (modulo the uncertainty in the calculation of $m_h$
for which FeynHiggs \cite{FeynHiggs} was used here).

\begin{figure}[htb!]
\begin{center}    
\epsfxsize = 0.5\textwidth
\epsffile{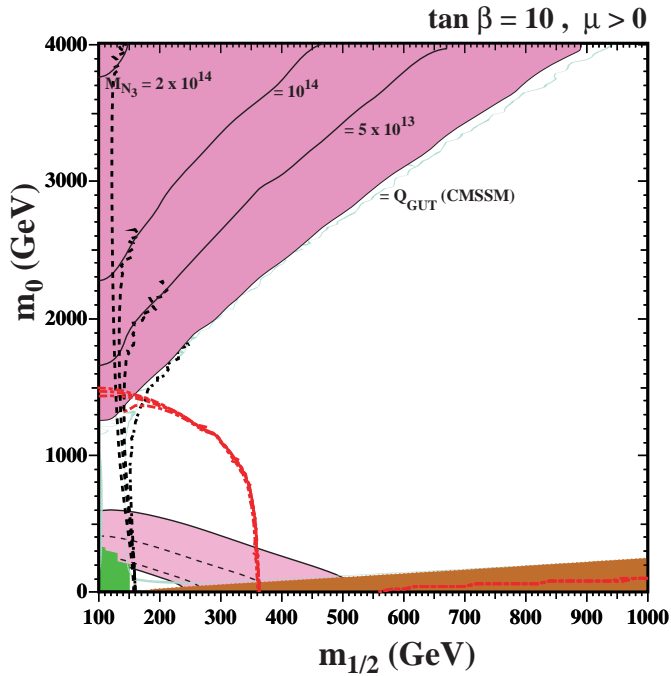 }
\end{center}        
\caption{The ($m_0,m_{1/2}$) plane for $\tan \beta = 10$, $A_0 = 0$, $m_{\nu_3}=0.05$eV, 
and four choices of $M_{N_3}$. The region labelled $Q_{\rm GUT}$ corresponds to the CMSSM.
The black solid lines represent the boundaries of the regions above which (i.e. for bigger $m_0$) there is no electroweak symmetry breaking and $\mu^2<0$. 
Contours and shaded regions are described in the text. 
}
\label{focusm0m12}
\end{figure}

Our qualitative discussion so far can be summarized in Fig. \ref{focusm0m12} 
which shows the ($m_0,m_{1/2}$) plane for $\tan \beta = 10$ and $A_0 = 0$. 
In the lower right corner of the figure, the brown shaded region corresponds to 
parameters for which the LSP is the charged partner of the tau lepton and as such is excluded.
Running along this region is the so-called WMAP co-annihilation strip, 
where co-annihilations between neutralinos and staus are largely responsible for
obtaining the WMAP relic density. This area (as in the previous figure) is shaded turquoise. 
In the lower left corner of the figure, the small dark (green) shaded region is excluded as the 
supersymmetric contributions to $b \to s \gamma$ disagree with the experimental determination
\cite{bsgex}, while the light pink shaded region is favored by the measurement of the muon 
anomalous magnetic moment at the 2-$\sigma$ level~\cite{g-2}\footnote{We have used
$\delta a_\mu = 24.6 \pm 8.0$ for the discrepancy between theory and experiment based
on $e^+e^-$ data \cite{davier}.}. The region between the pair of dashed line gives agreement
to within $1 \sigma$ of the experimental result.
The red dot-dashed contour corresponds to a
Higgs mass of 114.4 GeV.  At lower $m_{1/2}$, the Higgs boson would be lighter, which is 
excluded by its non-observation at LEP~\cite{LEPsusy}.  We also plot a black dashed 
contour for $m_{\chi^{\pm}}=104$ GeV, the region at lower $m_{1/2}$ also being 
excluded by LEP. These contours are shown for four choices of $M_{N_3}$ as labelled. 
Note that the stau co-annihilation strip is barely sensitive to the choices of $M_{N_3}$ made here
as are other quantities at low $m_0$.
This is reasonable because the terms involving a neutrino Yukawa coupling in the RGE of the slepton doublets can become large when the scalar masses or trilinear A terms are large as seen in Eq. (\ref{mhurge}). Because the gaugino masses and the right-handed stau are not affected by a neutrino Yukawa coupling at the one loop-level, the ratio of the stau and gaugino mass would not be affected significantly by the right-handed neutrino when $m_0$ and $A$ are small.

In the upper left corner of  Fig. \ref{focusm0m12}, we see the (pink) shaded region
where there is no radiative electroweak symmetry breaking.  Along the boundary of this region,
we find the focus point strip where again the relic density of neutralinos is within the WMAP range.
The shaded region corresponds to the standard CMSSM model or equivalently
the $\nu$CMSSM with $M_{N_3} = Q_{\rm GUT} \simeq 2 \times 10^{16}$ GeV.
Within the shaded region, we also show how the region of no
radiative electroweak symmetry breaking recedes to higher $m_0$ for values of $M_{N_3} <
Q_{\rm GUT}$. The boundary in each case is shown labelled by the chosen value of $M_{N_3}$.
The WMAP focus point strip would run along each boundary. 
Note that for very low $M_{N_3}$ ($\sim 2 \times 10^{12}$ GeV), we effectively recover
the CMSSM as the neutrino Yukawa couplings are too small to have any effect on RGEs.
As $M_{N_3}$ is increased, we see the disappearance of the focus point region as discussed above.

\section{Higgs funnel region}
\label{higfun}

\begin{figure}[ht!]
\begin{center}    
\epsfxsize = 0.5\textwidth
\epsffile{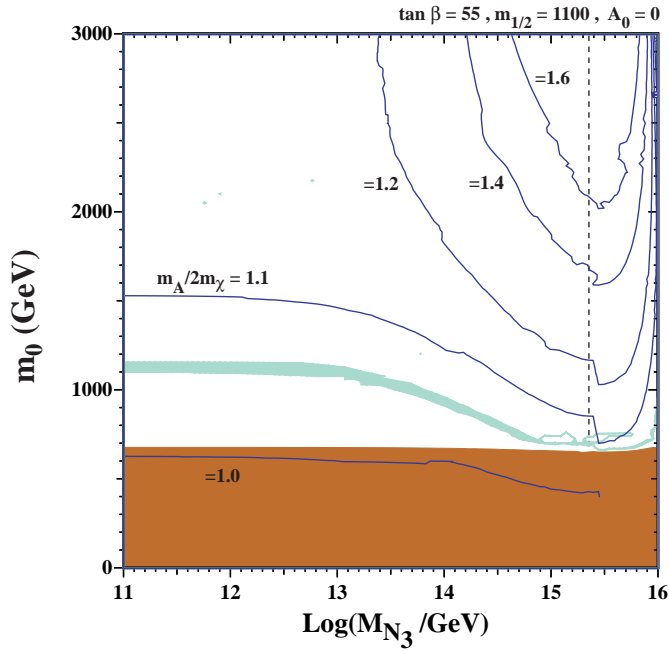}
\end{center}        
\caption{
\label{funnelMnm0}
The ($m_0,M_{N_3}$) plane for fixed 
$m_{1/2}=1100$ GeV, $\tan \beta =55, A_0=0$, and $m_{\nu_3}=0.05$ eV.
The funnel region (turquoise) 
compatible with Eq. (\ref{wmap}) runs almost horizontally 
around $m_0 \simeq 1100$ GeV. Contours of constant $m_A/2m_\chi$ are also shown.
To the right of the vertical dashed line, ${y^2_{N_3}}/{16\pi^2}>1/10$ and our perturbative
expansion becomes suspect.}
\end{figure}

\begin{figure}[ht]
\begin{center}    
\epsfxsize = 0.5\textwidth
\epsffile{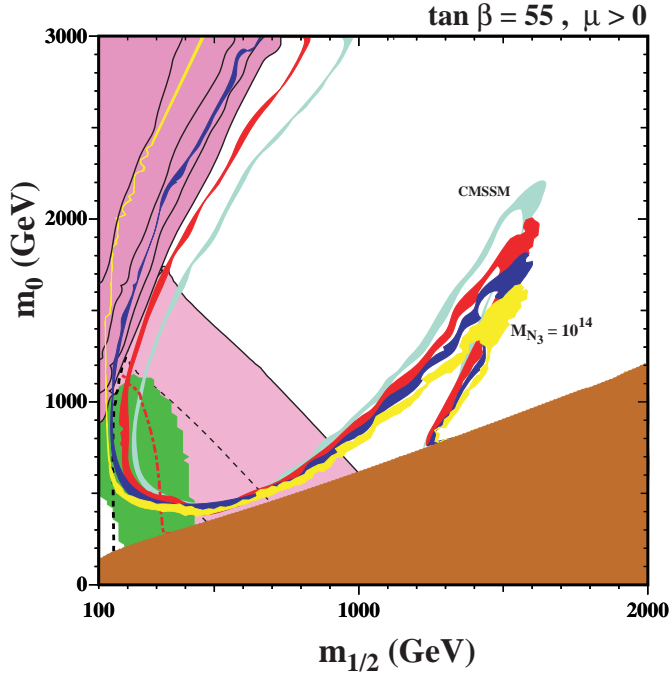}
\end{center}        
\caption{
\label{funfig}
As in Fig. \ref{focusm0m12},
the funnel regions for the CMSSM and for different values of $M_{N_3}=2 \times 10^{13}, 5\times 10^{13},10^{14}$ GeV in the $\nu$CMSSM. }
\end{figure}

Another efficient neutralino annihilation process which results in the requisite 
thermal relic density is through the s-channel Higgs (A and H) resonances and is possible 
at large $\tan \beta$. The dominant  
contributions are through the pseudo-scalar A which is CP odd and can couple to initial s-wave states (while the couplings through the CP-even H suffer from p-wave suppressions due to the small velocity of the neutralinos), with the dominant final states $b\bar{b}$ whose coupling is enhanced by a factor $\propto m_b \tan \beta$.
At large $\tan \beta$, the resonances become broad enough 
(typically $10\sim 50$GeV) so that sufficient neutralino annihilations can occur 
even if $m_{\chi}$ is several partial widths away from the exact resonance region 
$2m_{\chi}= m_A$.
The change in the RG evolution of $m^2_{H_u}$ in the $\nu$CMSSM also affects the A-pole funnel regions.   This can be understood from 
the tree-level Higgs potential which gives 
\ba
\label{mA}
m_A^2=m_{H_d}^2+m^2_{H_u}+2|\mu|^2
\ea
which, for  moderate to large $\tan \beta$, becomes
\ba
m_A^2\approx m_{H_d}^2-m^2_{H_u}-m_Z^2 .
\ea
For fixed $m_0$, and noting that the gaugino mass RGEs include $y_N$ at the two-loop level, we 
would expect the Higgs funnel regions to move towards larger $m_{1/2}$ for larger $y_N$.
This is because $y_N$ increases $m_A$ by making 
$m_{H_u}^2$ more negative and hence requires larger $m_{1/2}$ in order 
to satisfy $2 m_{\chi}\sim m_A$.
For $m_{1/2}$ fixed, the increase in $m_A$ from an increase in $y_N$ must
be compensated by a decease in $m_0$.
This is because a decrease in $m_0$ can change $m_{H_d}^2$ (which is affected by $y_N$ at the two-loop level) to cancel the change in $m_{H_u}^2$ and can also suppress the terms in the RGE which involve the products of $y_N^2$ and scalar mass squared. These effects are summarized in Figs. \ref{funnelMnm0} and \ref{funfig}. Fig. \ref{funnelMnm0} shows the 
contours of $m_A/2m_{\chi}$ in the $(m_0,M_{N_3})$ plane. Here, 
we can see the broad Higgs funnel region
extending between the $m_A/2m_{\chi} = 1$ and 1.1 contours. 
The Higgs resonance is so efficient
such that the WMAP strips typically show up slightly off the
exact resonance $m_{A}/2m_{\chi}=1$ contour line. In the brown shaded region at lower $m_0$, the lightest stau is the LSP and hence this
region is excluded.  Between this region and the WMAP strip, the relic density is lower
than the WMAP range.

Fig. \ref{funfig} shows the ($m_0,m_{1/2}$) plane at $\tan \beta =55$ for which the funnel region
is clearly visible. As in Fig. \ref{focusm0m12}, we show the position of the
Higgs and chargino mass contours at the LEP limit. 
Also shown are the shaded regions excluded by a charged stau LSP and $b \to s \gamma$
as well as the shaded region preferred by $g_\mu -2$.  The shaded region in the upper
left again corresponds to the region with no radiative electroweak symmetry breaking in the CMSSM.
The shaded region labelled CMSSM, corresponds to the funnel region where the 
relic density agrees with the WMAP range.
The remaining funnel-like shaded regions correspond to the shift in the funnel  
in the $\nu$CMSSM for different values of 
$M_{N_3} = 2 \times 10^{13}$ GeV, $5 \times 10^{13}$ GeV and $10^{14}$ GeV. 
Fig. \ref{funnelMnm0} corresponded to the choice $m_{1/2} = 1100$ GeV, and 
as one can see for values of $m_0$ above the line of neutralino-stau degeneracy and below 
the WMAP funnel,
s-channel neutralino annihilations via the heavy Higgs 
are so strong near $m_{\tilde{\tau}}\sim m_{\chi}$ 
that the relic density is pushed to values far below the WMAP
range.  As such, for this value of $m_{1/2}$ there is no WMAP stau coannihilation region.

As the neutrino Yukawa coupling becomes larger, the funnel regions move down towards the stau coannihilation regions. Also note that the larger values of $\mu$ obtained through 
the effects of the neutrino Yukawa couplings increase the left-right mixings in the stau mass matrix, especially for  large $\tan \beta$, due to the off-diagonal term $-m_{\tau}(\mu \tan \beta +A_{\tau})$. The neutrino Yukawa couplings can also 
increase the left-handed component in the lighter slepton mass eigenstate by suppressing the left-handed slepton mass 
as discussed in \S \ref{sub2}.

The focus point strip is also visible in Fig. \ref{funfig}.  For the case of the CMSSM, in the area 
between the focus point strip and the shaded region corresponding to $\mu^2 < 0$,
the lightest neutralino acquires a significant Higgsino component and the relic
density is below the WMAP range. As discussed in the previous section,
as $M_{N_3}$ is increased, the focus point strips move to larger values of $m_0$.
The black solid lines to the left of each colored strip correspond to the edge of the
region where radiative electroweak symmetry breaking is lost.

\section{Effects from the first two generation right-handed neutrinos}
\label{invert}

We have so far discussed the effects of only one heavy right-handed neutrino 
in the third generation
exemplifying a normal hierarchy scenario $m_{\nu 3}\gg 
m_{\nu 1}\approx m_{\nu 2} $. That is we have implicitly assumed that the Yukawa couplings
for the first two generation neutrinos are small and do not affect the sparticle spectrum.
We next consider the effects of multiple right-handed neutrinos in other generations for comparison. We then go on to explore how the parameter space at relatively low $m_{1/2}$ and $m_0$ is affected by the seesaw mechanism, with an emphasis on the 
emergence of sneutrino coannihilation regions.

\subsection{Focus and Higgs funnel regions}
\label{sub1}

We now consider a model which includes $N_1$ and $N_2$ with equal masses,  $M_{N_1}=M_{N_2}$, corresponding to an 
inverted mass hierarchy  $m_{\nu 3}\ll m_{\nu 1}= m_{\nu 2}$. 
Even though this toy model is not complete in that we ignore the neutrino mixings for simplicity, it does illustrate the features of the 
$e$ and $\mu$ right-handed neutrinos which are distinct from those of the $\tau$ right-handed neutrino and those of the CMSSM. We save a more complete analysis including all three neutrinos and mixings
for a future publication.

\begin{figure}[htb!]
         \begin{center}
                  \includegraphics[width=0.47\textwidth]{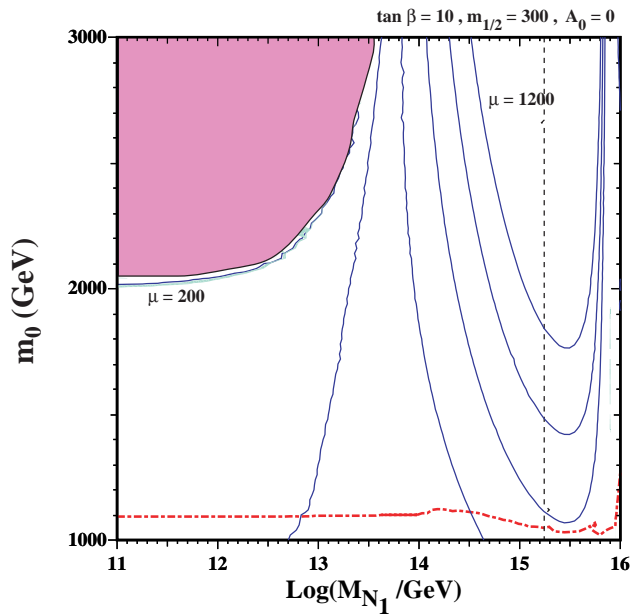}
            \end{center}
\caption{As in Fig. \ref{focusMnm0}, the ($m_0,M_{N_1}$) plane for fixed 
$m_{1/2}=300$ GeV, $\tan \beta =10, A_0=0$, and $m_{\nu_1}=m_{\nu_2}=0.05$ eV.
It is assumed that $M_{N_1} = M_{N_2}$.
}
     \label{i=3} 
     \end{figure}

The effects of $N_1$ and $N_2$ on the Higgs masses are 
qualitatively analogous to those of $N_3$ because the 
neutrino Yukawa coupling term, $y_{ij}N_i L_j H_u$, ($y_{ij}$ is diagonal in this example) which gives the key interactions between $N_i$ and the Higgs bosons
has a common form for each generation. 
Thus we expect similar effects on the RG evolution of $m^2_{H_u}$ (or $\mu$) as discussed above.
The effect of making $m_{H_u}^2$ even more negative is, however, bigger now because 
there are contributions from both first and second generations, and hence the shift and disappearance of the focus point and funnel regions are more prominent quantitatively. 
As one can see from Fig. \ref{i=3}, the edge of the region with no electroweak symmetry breaking
at large $m_0$ is shifted down in $M_{N_1}$ by a factor of two relative to where it was in
Fig. \ref{focusMnm0} for $N_3$.
The sensitivity of 
$\mu$ on $M_{N_1},M_{N_2}$ is also shown in Fig. \ref{i=3}. 

\subsection{Sneutrino coannihilation regions}
\label{sub2}

Our final consideration is the sneutrino coannihilation region at relatively 
small $m_0$ and $m_{1/2}$. Here,  the qualitative interpretation of the 
parameter space compatible with the observed dark matter abundance can differ significantly between the two toy models discussed in this paper.
Sneutrino coannihilation regions where the sneutrino is the NLSP with a neutralino LSP 
are not realized in the CMSSM because  the
right-handed species are usually lighter than the 
corresponding left-handed slepton doublets. This is because the
SU(2) gauge coupling terms in the RGEs tend to push the 
masses of slepton doublets up over their right-handed counter parts when universal boundary conditions at the high energy scale are chosen as in the CMSSM. In the $\nu$CMSSM, however, the neutrino Yukawa coupling which involves the left-handed 
SU(2) doublet $y_N N L H_u$ can push the left-handed slepton doublet mass down while running 
from $Q=Q_{GUT}$ down to $Q=M_N$, enabling the slepton doublets to be lighter than their corresponding right-handed singlets at the electroweak scale.

In the model with a single right handed neutrino ($N_3$), 
the realization of the sneutrino coannihilation regions tends to require a large value of  $A_0$ and a moderate value of $\tan \beta$ \cite{kov}. 
A small universal scalar mass is generally required to make the sneutrino light, which results in the necessity for large $A_0$ because the effects of the neutrino Yukawa coupling show up in the 
RGEs in the form of a 
product between $y_N$ and the sum of scalar masses and $A_0$ as given in Eq. (\ref{mhurge}). 
Large $A_0$ also helps enhance the 
stop loop contributions to the Higgs mass which relaxes the tight constraint from Higgs mass lower bound when $m_{1/2}$ is small
(low $m_{1/2}$ is also preferred in order  to obtain a light sneutrino by suppressing the 
the RG evolution of the slepton doublet due to gauge interactions). 
Moderate $\tan \beta$ is preferred because 
at large $\tan \beta$  the induced mixing in the stau mass matrix is too large and causes one of  
the stau mass eigenstates to be run below the
sneutrino mass.

\begin{figure}[htb!]
             \centering
                 \includegraphics[width=0.47\textwidth]{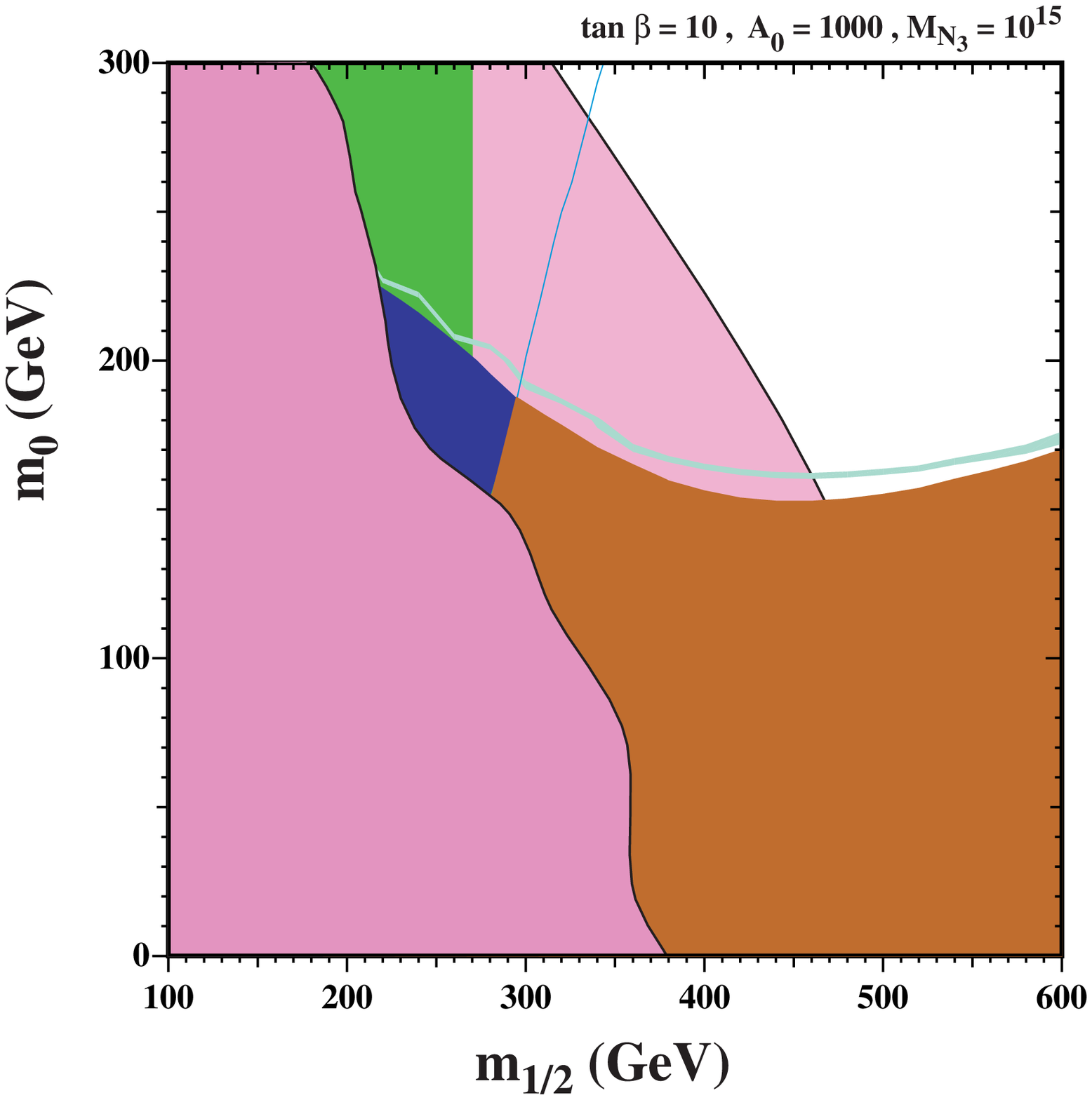}
                 \includegraphics[width=0.47\textwidth]{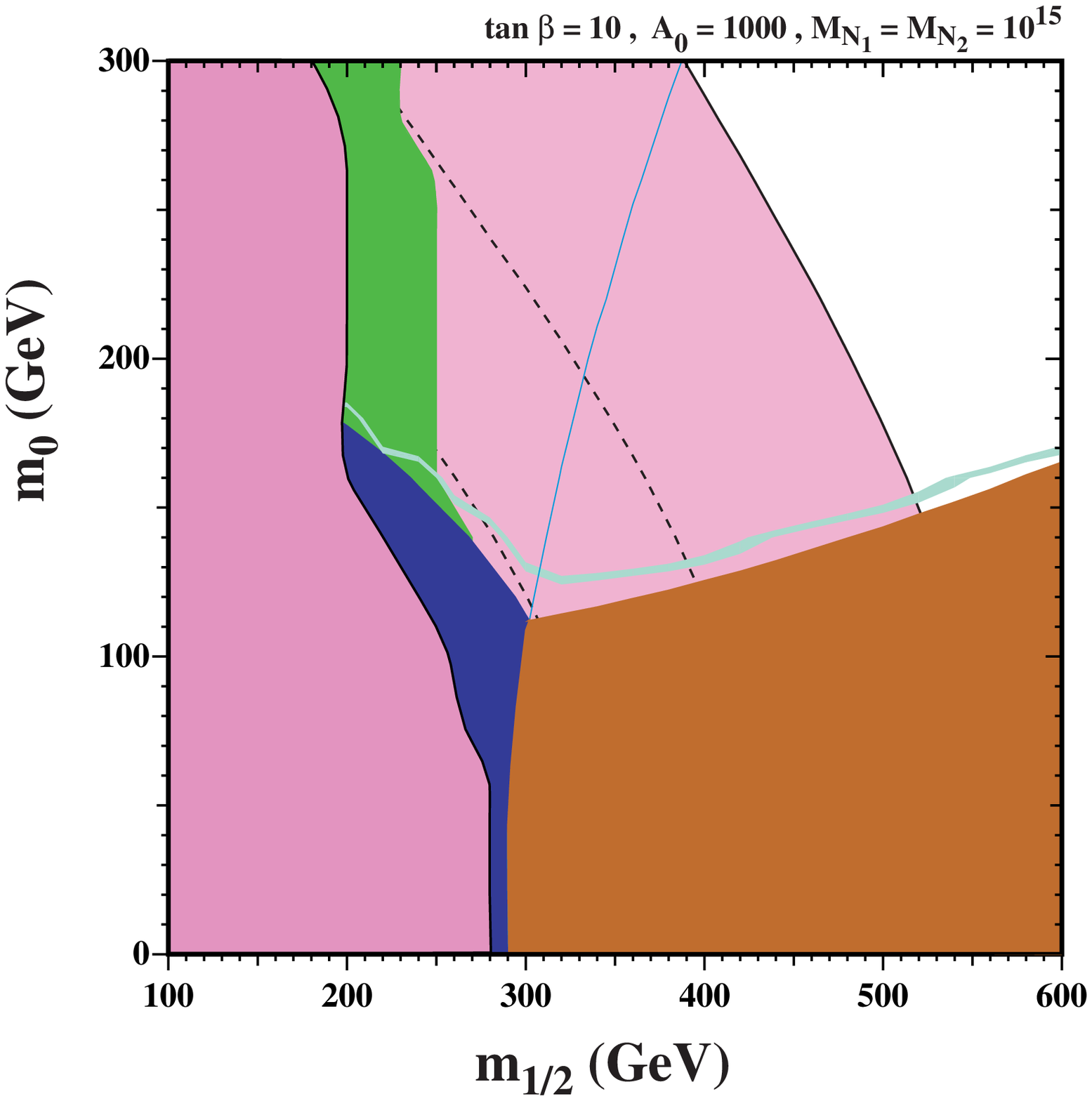}
 \caption{The $(m_0,m_{1/2})$ plane in the $\nu$CMSSM. In both panels, 
 $\tan \beta = 10$ and $A_0 = 1000$ GeV. In the left panel, $M_{N_3} = 10^{15}$ GeV
 and corresponds to the case of a normal neutrino mass hierarchy while in the 
 right panel, $M_{N_1} = M_{N_2} = 10^{15}$ GeV corresponding to an inverted 
 hierarchy. Contours and shading are described in the text.
 }
     \label{selsmu} 
     \end{figure}

Fig. \ref{selsmu} shows the $(m_0,m_{1/2})$ plane in the $\nu$CMSSM for the two types of models
considered (corresponding to the normal and inverted hierarchies).
As in Figs. \ref{focusm0m12} and \ref{funfig}, the brown shaded region is excluded because
the LSP is the charged partner of the stau.  In the left portion of each plane, there are
areas where one or more of the sparticles are tachyonic.  This relatively large region is shaded pink.
There is also a region which is excluded by measurements of $b \to s \gamma$ and these (shaded green) also exclude low values of $m_{1/2}$. In the background of each panel, there is a light
pink shaded region where the value of $g_\mu - 2$ is in agreement with the observed discrepancy
to within $2 \sigma$.  In the right panel, this region is more prominent and the area within
the dashed contour satisfies the $g_\mu -2 $ constraint within 1 $\sigma$. The dark blue shaded region
in each figure is a direct result of the $\nu$CMSSM and corresponds to the area where 
a sneutrino is the LSP.  This region is also excluded \cite{directleft}.  Finally, there is the region where the relic density matches the WMAP determination as shown by the thin turquoise strips
which track either the stau or sneutrino LSP regions. The relic density in each case is
brought to an acceptable value through neutralino coannihilations with staus 
\cite{stau} and/or sneutrinos \cite{john2}.

For our normal hierarchy model shown in the left panel of Fig. \ref{selsmu}, the stau coannihilation region is adjacent to the tau sneutrino NLSP regions (the thin blue line shows the position of the 
contour $m_{\tilde{\nu}}/m_{\tilde \tau}=1$). Analogous to the $\tau$ sneutrino coannihilation regions,  $e/\mu$ sneutrino coannihilation regions appear in our second toy model shown in 
the right panel of Fig. \ref{selsmu}.  In this case, the effects of the right-handed neutrinos in 
reducing the left-handed slepton mass are more prominent than in the model with a normal 
neutrino mass hierarchy. For instance, 
sneutrino coannihilation regions show up for $A_0\gtrsim 700$GeV for our 
inverted mass hierarchy toy model with $M_{N_1}=M_{N_2}=10^{15}$GeV, $m_{\nu_1}=m_{\nu_2}=0.05$eV, while $A_0\gtrsim 1000$GeV is required for our normal mass hierarchy toy model with $M_{N_3}=10^{15}$GeV, $m_{\nu_3}=0.05$eV (and other parameter values are fixed as in Fig. \ref{selsmu}).
For the electron/muon sneutrino coannihilation regions, the WMAP strip transition corresponds to the abrupt change from the mostly horizontal right-handed stau NLSP region to the more vertical left-handed electron/muon sneutrino NLSP region. For the tau sneutrino NLSP, in contrast, the transition from the stau coannihilation region to the tau sneutrino coannihilation region corresponds to the smooth change of the stau chirality which is initially right-chiral dominated with an increasing left-chiral component for smaller $m_{1/2}$, before it eventually 
shifts to the tau sneutrino NLSP region. This is expected as the left-handed slepton doublet becomes light due to a large neutrino Yukawa coupling in a region with small $m_{1/2}$. The tau sneutrino typically accompanies an almost degenerate left-handed stau 
state with their small mass splitting coming from D-term contributions, hence resulting in nearly 
parallel contours 
for $m_{\tilde{\nu}_{\tau}}/m_{\chi}=1$ and $m_{\tilde{\tau}}/m_{\chi}=1$ in the left panel in Fig. \ref{selsmu}, while a more abrupt transition from the 
(right-handed) stau coannihilation to the (left-handed) electron/muon sneutrino coannihilation regions arises in the right panel.

\section{Conclusion}
\label{conc}
The CMSSM has been studied extensively as a template to see the general features of a simple supersymmetric model.  Given the necessity for neutrino masses, the CMSSM implicitly assumes
that the additional operators relevant to neutrino masses are sufficiently 
small that their presence in the running 
of the supersymmetric particle spectrum can be ignored. In effect, from the viewpoint of the 
seesaw scheme, such an extension of the CMSSM assumes a relatively
low value for the heavy right-handed neutrino masses.
The $\nu$CMSSM is an extension of the CMSSM and explicitly includes three right-handed neutrinos 
in a general type-I seesaw setup while keeping the universal boundary conditions as in CMSSM.  
Here, we have studied how the dark matter abundances are affected by presence of right-handed neutrinos whose masses are close to the GUT scale.
Though we ignored flavor mixing and the CP violating phase in the neutrino mass matrix,  our two simple models illustrate the intriguing effects of GUT scale right-handed neutrinos.  These include the change in the focus point scale, shifts in the funnel region and the realization of light left-handed slepton doublets. Clearly the GUT-scale seesaw mechanism and its effects on low-energy observables deserve further study. 

This work was supported by Michigan Center for Theoretical Physics (KK), DOE grant DE-FG02-94ER-40823 and the William I. Fine 
Theoretical Physics Institute (KAO).



\begin{thebibliography}{99}


\bibitem{seesaw}
P.~Minkowski, Phys.\ Lett.\ B {\bf 67} (1977) 421.; 
M. Gell-Mann, P. Ramond, and R. Slansky, in  
$\underline{Supergravity}$, eds. D.Z. Freedman and P. van  
Nieuwenhuizen, North Holland (1979); 
T. Yanagida, in Proceedings of the Workshop on  
the Unified Theory and The Baryon Number of the Universe, eds O. Sawada  
and S. Sugamoto. KEK79-18 (1979). 



\bibitem{borzu}
  F.~Borzumati and A.~Masiero,
  Phys.\ Rev.\ Lett.\  {\bf 57}, 961 (1986).



\bibitem{hisa}
  J.~Hisano, T.~Moroi, K.~Tobe and M.~Yamaguchi,
  Phys.\ Rev.\  D {\bf 53}, 2442 (1996)
  [arXiv:hep-ph/9510309].
  
\bibitem{casa3}
 J.~A.~Casas, J.~R.~Espinosa, A.~Ibarra and I.~Navarro,
 Nucl.\ Phys.\  B {\bf 569}, 82 (2000)
 [arXiv:hep-ph/9905381].
 
\bibitem{Lavignac:2001vp}
 S.~Lavignac, I.~Masina and C.~A.~Savoy,
models,''
 Phys.\ Lett.\  B {\bf 520}, 269 (2001)
 [arXiv:hep-ph/0106245].
 
\bibitem{Ellis:2001xt}
 J.~R.~Ellis, J.~Hisano, S.~Lola and M.~Raidal,
 Nucl.\ Phys.\  B {\bf 621}, 208 (2002)
 [arXiv:hep-ph/0109125].


  \bibitem{bm}
B.~A.~Campbell and D.~W.~Maybury,
  JHEP {\bf 0704}, 077 (2007)
  [arXiv:hep-ph/0603053].

\bibitem{de}
  A.~Dedes, H.~E.~Haber and J.~Rosiek,
  JHEP {\bf 0711}, 059 (2007)
  [arXiv:0707.3718 [hep-ph]].


\bibitem{iba}
  A.~Ibarra and C.~Simonetto,
  JHEP {\bf 0804}, 102 (2008)
  [arXiv:0802.3858 [hep-ph]].


\bibitem{hir}
  M.~Hirsch, J.~W.~F.~Valle, W.~Porod, J.~C.~Romao and A.~Villanova del Moral,
  Phys.\ Rev.\  D {\bf 78}, 013006 (2008)
  [arXiv:0804.4072 [hep-ph]].





\bibitem{casa}
  J.~A.~Casas, J.~R.~Espinosa, A.~Ibarra and I.~Navarro,
  Phys.\ Rev.\  D {\bf 63}, 097302 (2001)
  [arXiv:hep-ph/0004166].
  
\bibitem{how4}
  H.~Baer, C.~Balazs, J.~K.~Mizukoshi and X.~Tata,
  Phys.\ Rev.\  D {\bf 63}, 055011 (2001)
  [arXiv:hep-ph/0010068].

\bibitem{baer2}
  H.~Baer, C.~Balazs, M.~Brhlik, P.~Mercadante, X.~Tata and Y.~Wang,
  Phys.\ Rev.\  D {\bf 64}, 015002 (2001)
  [arXiv:hep-ph/0102156].

\bibitem{bla}
  G.~A.~Blair, W.~Porod and P.~M.~Zerwas,
  Eur.\ Phys.\ J.\  C {\bf 27}, 263 (2003)
  [arXiv:hep-ph/0210058].
  

 
\bibitem{hitoshi2}
  M.~R.~Buckley and H.~Murayama,
  Phys.\ Rev.\ Lett.\  {\bf 97}, 231801 (2006)
  [arXiv:hep-ph/0606088].

\bibitem{gordy3}
  G.~L.~Kane, P.~Kumar, D.~E.~Morrissey and M.~Toharia,
  Phys.\ Rev.\  D {\bf 75}, 115018 (2007)
  [arXiv:hep-ph/0612287].

  
\bibitem{hints}
  J.~A.~Casas, A.~Ibarra and F.~Jimenez-Alburquerque,
  JHEP {\bf 0704}, 064 (2007)
  [arXiv:hep-ph/0612289].



\bibitem{kov}
  K.~Kadota, K.~A.~Olive and L.~Velasco-Sevilla,
  Phys.\ Rev.\  D {\bf 79}, 055018 (2009)
  arXiv:0902.2510 [hep-ph].








\bibitem{petcov}
  S.~T.~Petcov, S.~Profumo, Y.~Takanishi and C.~E.~Yaguna,
  Nucl.\ Phys.\  B {\bf 676}, 453 (2004)
  [arXiv:hep-ph/0306195].


\bibitem{cali}
  L.~Calibbi, Y.~Mambrini and S.~K.~Vempati,
  JHEP {\bf 0709}, 081 (2007)
  [arXiv:0704.3518 [hep-ph]].

\bibitem{barg}
  V.~Barger, D.~Marfatia and A.~Mustafayev,
  Phys.\ Lett.\  B {\bf 665}, 242 (2008)
  [arXiv:0804.3601 [hep-ph]].


\bibitem{gom}
  M.~E.~Gomez, S.~Lola, P.~Naranjo and J.~Rodriguez-Quintero,
  arXiv:0901.4013 [hep-ph].


\bibitem{est}
  J.~N.~Esteves, M.~Hirsch, S.~Kaneko, W.~Porod and J.~C.~Romao,
  arXiv:0907.5090 [hep-ph].


\bibitem{barg2}
  V.~Barger, D.~Marfatia, A.~Mustafayev and A.~Soleimani,
  arXiv:0908.0941 [hep-ph].








\bibitem{funnel}
M.~Drees and M.~M.~Nojiri,
Phys.\ Rev.\ D {\bf 47} (1993) 376 [arXiv:hep-ph/9207234];
H.~Baer and M.~Brhlik,
Phys.\ Rev.\ D {\bf 53} (1996) 597 [arXiv:hep-ph/9508321];
  Phys.\ Rev.\  D {\bf 57} (1998) 567
  [arXiv:hep-ph/9706509];
H.~Baer, M.~Brhlik, M.~A.~Diaz, J.~Ferrandis, P.~Mercadante, P.~Quintana and X.~Tata,
  Phys.\ Rev.\  D {\bf 63} (2000) 015007
  [arXiv:hep-ph/0005027];
 A.~B.~Lahanas, D.~V.~Nanopoulos and V.~C.~Spanos,
  Mod.\ Phys.\ Lett.\  A {\bf 16} (2001) 1229
  [arXiv:hep-ph/0009065].

\bibitem{cmssm}
J.~R.~Ellis, T.~Falk, K.~A.~Olive and M.~Schmitt,
Phys.\ Lett.\ B {\bf 388} (1996) 97
[arXiv:hep-ph/9607292];
Phys.\ Lett.\ B {\bf 413} (1997) 355
[arXiv:hep-ph/9705444];
J.~R.~Ellis, T.~Falk, G.~Ganis, K.~A.~Olive and M.~Schmitt,
Phys.\ Rev.\ D {\bf 58} (1998) 095002
[arXiv:hep-ph/9801445];
V.~D.~Barger and C.~Kao,
Phys.\ Rev.\ D {\bf 57} (1998) 3131
[arXiv:hep-ph/9704403];
J.~R.~Ellis, T.~Falk, G.~Ganis and K.~A.~Olive,
Phys.\ Rev.\ D {\bf 62} (2000) 075010
[arXiv:hep-ph/0004169];
V.~D.~Barger and C.~Kao,
Phys.\ Lett.\ B {\bf 518} (2001) 117
[arXiv:hep-ph/0106189];
L.~Roszkowski, R.~Ruiz de Austri and T.~Nihei,
JHEP {\bf 0108} (2001) 024
[arXiv:hep-ph/0106334];
A.~B.~Lahanas and V.~C.~Spanos,
Eur.\ Phys.\ J.\ C {\bf 23} (2002) 185
[arXiv:hep-ph/0106345];
A.~Djouadi, M.~Drees and J.~L.~Kneur,
JHEP {\bf 0108} (2001) 055
[arXiv:hep-ph/0107316];
U.~Chattopadhyay, A.~Corsetti and P.~Nath,
Phys.\ Rev.\ D {\bf 66} (2002) 035003
[arXiv:hep-ph/0201001];
J.~R.~Ellis, K.~A.~Olive and Y.~Santoso,
New Jour.\ Phys.\  {\bf 4} (2002) 32
[arXiv:hep-ph/0202110];
H.~Baer, C.~Balazs, A.~Belyaev, J.~K.~Mizukoshi, X.~Tata and Y.~Wang,
JHEP {\bf 0207} (2002) 050
[arXiv:hep-ph/0205325];
R.~Arnowitt and B.~Dutta,
arXiv:hep-ph/0211417.

\bibitem{efgosi}
J.~R.~Ellis, T.~Falk, G.~Ganis, K.~A.~Olive and M.~Srednicki,
Phys.\ Lett.\ B {\bf 510} (2001) 236
[arXiv:hep-ph/0102098].


\bibitem{analytic}
  J.~L.~Feng and K.~T.~Matchev,
  Phys.\ Rev.\  D {\bf 63}, 095003 (2001)
  [arXiv:hep-ph/0011356];
  J.~L.~Feng, K.~T.~Matchev and T.~Moroi,
  Phys.\ Rev.\  D {\bf 61}, 075005 (2000)
  [arXiv:hep-ph/9909334].


\bibitem{dm}
  J.~L.~Feng, K.~T.~Matchev and F.~Wilczek,
  Phys.\ Lett.\  B {\bf 482}, 388 (2000)
  [arXiv:hep-ph/0004043].


\bibitem{eoss}
J.~R.~Ellis, K.~A.~Olive, Y.~Santoso and V.~C.~Spanos,
Phys.\ Lett.\ B {\bf 565} (2003) 176
[arXiv:hep-ph/0303043].



\bibitem{cmssmwmap}
H.~Baer and C.~Balazs,
  JCAP {\bf 0305}, 006 (2003)
  [arXiv:hep-ph/0303114];
A.~B.~Lahanas and D.~V.~Nanopoulos,
  Phys.\ Lett.\  B {\bf 568}, 55 (2003)
  [arXiv:hep-ph/0303130];
U.~Chattopadhyay, A.~Corsetti and P.~Nath,
  Phys.\ Rev.\  D {\bf 68}, 035005 (2003)
  [arXiv:hep-ph/0303201];
   C.~Munoz,
  Int.\ J.\ Mod.\ Phys.\  A {\bf 19}, 3093 (2004)
  [arXiv:hep-ph/0309346].




\bibitem{hb}
  K.~L.~Chan, U.~Chattopadhyay and P.~Nath,
  Phys.\ Rev.\  D {\bf 58}, 096004 (1998)
  [arXiv:hep-ph/9710473];
J.~L.~Feng, K.~T.~Matchev and T.~Moroi,
  Phys.\ Rev.\ Lett.\  {\bf 84}, 2322 (2000)
  [arXiv:hep-ph/9908309]; 
  H.~Baer, T.~Krupovnickas and X.~Tata,
  JHEP {\bf 0307}, 020 (2003)
  [arXiv:hep-ph/0305325];
  G.~Belanger, S.~Kraml and A.~Pukhov,
  Phys.\ Rev.\  D {\bf 72}, 015003 (2005)
  [arXiv:hep-ph/0502079];
  H.~Baer, T.~Krupovnickas, S.~Profumo and P.~Ullio,
  JHEP {\bf 0510} (2005) 020
  [arXiv:hep-ph/0507282].




\bibitem{wmap}
 J.~Dunkley {\it et al.}  [WMAP Collaboration],
  Astrophys.\ J.\ Suppl.\  {\bf 180}, 306 (2009)
  [arXiv:0803.0586 [astro-ph]].







\bibitem{ant}
  S.~Antusch and M.~Ratz,
  JHEP {\bf 0207}, 059 (2002)
  [arXiv:hep-ph/0203027].

\bibitem{ant2}
  S.~Antusch, J.~Kersten, M.~Lindner, M.~Ratz and M.~A.~Schmidt,
  JHEP {\bf 0503}, 024 (2005)
  [arXiv:hep-ph/0501272].



\bibitem{top}
    [Tevatron Electroweak Working Group and CDF Collaboration and D0 Collaboration],
  arXiv:0803.1683 [hep-ex], 
  arXiv:0903.2503 [hep-ex].

\bibitem{mtvary}
  J.~R.~Ellis, S.~Heinemeyer, K.~A.~Olive and G.~Weiglein,
  JHEP {\bf 0502}, 013 (2005)
  [arXiv:hep-ph/0411216].


\bibitem{focusvary}
  A.~Romanino and A.~Strumia,
  Phys.\ Lett.\  B {\bf 487}, 165 (2000)
  [arXiv:hep-ph/9912301];
 J.~R.~Ellis and K.~A.~Olive,
  Phys.\ Lett.\  B {\bf 514}, 114 (2001)
  [arXiv:hep-ph/0105004].

\bibitem{ytop}
  H.~Baer, T.~Krupovnickas and X.~Tata,
  JHEP {\bf 0307}, 020 (2003)
  [arXiv:hep-ph/0305325].



\bibitem{nuhm}
  H.~Baer, A.~Mustafayev, S.~Profumo, A.~Belyaev and X.~Tata,
  JHEP {\bf 0507}, 065 (2005)
  [arXiv:hep-ph/0504001].




\bibitem{nume}
  G.~Belanger, S.~Kraml and A.~Pukhov,
  Phys.\ Rev.\  D {\bf 72}, 015003 (2005)
  [arXiv:hep-ph/0502079];
B.~C.~Allanach, S.~Kraml and W.~Porod,
  JHEP {\bf 0303}, 016 (2003)
  [arXiv:hep-ph/0302102].

\bibitem{LEPsusy}
 R. Barate et al. (LEPH Collaboration, DELPHI Collaboration, L3 Collaboration, OPAL Collaboration, and The LEP Working Group for Higgs Boson Searches), Phys. Lett. B 565, 61 (2003); LEP Higgs Working Group, http://lephiggs.web.cern.ch/LEPHIGGS/Welcome.html; Joint LEP2 SUSY Working Group, https://lepsusy.web.cern.ch/lepsusy.








  
  \bibitem{FeynHiggs}
S.~Heinemeyer, W.~Hollik and G.~Weiglein,
{\it Comput.\ Phys.\ Commun.\ } {\bf 124} (2000) 76 
[arXiv:hep-ph/9812320];
S.~Heinemeyer, W.~Hollik and G.~Weiglein,
{\it Eur.\ Phys.\ J.\ C} {\bf 9} (1999) 343 
[arXiv:hep-ph/9812472];
 G.~Degrassi, S.~Heinemeyer, W.~Hollik, P.~Slavich and G.~Weiglein,
  Eur.\ Phys.\ J.\ C {\bf 28} (2003) 133
  [arXiv:hep-ph/0212020].;
  M.~Frank, T.~Hahn, S.~Heinemeyer, W.~Hollik, H.~Rzehak and G.~Weiglein,
  JHEP {\bf 0702} (2007) 047
  [arXiv:hep-ph/0611326].

  
\bibitem{bsgex}
  S.~Chen {\it et al.}  [CLEO Collaboration],
  Phys.\ Rev.\ Lett.\  {\bf 87} (2001) 251807
  [arXiv:hep-ex/0108032];
  P.~Koppenburg {\it et al.}  [Belle Collaboration],
  Phys.\ Rev.\ Lett.\  {\bf 93} (2004) 061803
  [arXiv:hep-ex/0403004].
  B.~Aubert {\it et al.}  [BaBar Collaboration],
  arXiv:hep-ex/0207076;
  E.~Barberio {\it et al.}  [Heavy Flavor Averaging Group (HFAG)],
  arXiv:hep-ex/0603003.

\bibitem{g-2} 
  G.~Bennett {\it et al.}\ [The Muon g-2 Collaboration],
  {\it Phys. Rev. Lett.} {\bf 92} (2004) 161802, 
  hep-ex/0401008;
  G.~Bennett {\it et al.}\ [The Muon g-2 Collaboration],
  {\em Phys.\ Rev.} {\bf D 73} (2006) 072003
  [arXiv:hep-ex/0602035].

\bibitem{davier}
  M.~Davier, A.~Hoecker, B.~Malaescu, C.~Z.~Yuan and Z.~Zhang,
  arXiv:0908.4300 [hep-ph].

  \bibitem{directleft}
  T.~Falk, K.~A.~Olive and M.~Srednicki,
  Phys.\ Lett.\  B {\bf 339}, 248 (1994)
  [arXiv:hep-ph/9409270];
C.~Arina and N.~Fornengo,
  JHEP {\bf 0711}, 029 (2007)
  [arXiv:0709.4477 [hep-ph]].
  
  \bibitem{stau}
  J.~R.~Ellis, T.~Falk and K.~A.~Olive,
  Phys.\ Lett.\  B {\bf 444}, 367 (1998)
  [arXiv:hep-ph/9810360];
  J.~R.~Ellis, T.~Falk, K.~A.~Olive and M.~Srednicki,
  Astropart.\ Phys.\  {\bf 13}, 181 (2000)
  [Erratum-ibid.\  {\bf 15}, 413 (2001)]
  [arXiv:hep-ph/9905481].

\bibitem{john2}
J.~R.~Ellis, K.~A.~Olive and Y.~Santoso,
  JHEP {\bf 0810}, 005 (2008)
  [arXiv:0807.3736 [hep-ph]].


\end{thebibliography}
\end{document}